\documentstyle[11pt]{article}
\begin{document}
\begin{titlepage}
\title{Relativistic formulation of quantum state diffusion?}
\author{Lajos Di\'osi
\thanks{E-mail: diosi.rmki.kfki.hu}\\
KFKI Research Institute for Particle and Nuclear Physics\\
H-1525 Budapest 114., P.O.Box 49, Hungary\\\\
{\it e-print archives ref.: quant-ph/9803039}\hfill}
\date{March 12, 1998}
\maketitle
\begin{abstract}
The recently reported relativistic formulation of the well-known 
non-relativistic quantum state diffusion is seriously mistaken.
It predicts, for instance, inconsistent measurement outcomes
for the same system when seen by two different inertial observers.
\end{abstract}
\end{titlepage}

{\it Introduction.}
Breuer and Petruccione (BP) \cite{BP} have quite recently reported the
construction of a relativistic generalization of the stochastic 
Ito-Schr\"odinger-equations now widely used for open quantum systems.
I have found that the model is completely mistaken. It can not  
reproduce even the ordinary quantum state diffusion theory as its 
non-relativistic limit. Its non-relativistic limit predicts, 
for instance, inconsistent experimental results for slowly moving 
observers. The reader might ignore my own interpretation of BP's 
{\it Concept} and might read directly the {\it Counter-example}. 

{\it The Concept.}
In Dirac's electron theory, to each space-like hyper-plane $\sigma$ 
of the Minkowski-space a quantum state $\psi$ is attributed. 
Such $\psi$ is interpreted as
the quantum state which is seen by the inertial observer residing
on the hyper-plane $\sigma$. Let, for concreteness, the hyper-planes be
parameterized by their unit normal vectors $n$ and by their distance 
$a$ from the origin. If we vary the hyper-surface, 
the state vector $\psi(\sigma)\equiv\psi(n,a)$ transforms unitarily:
$$
d\psi=-iaH\psi -i dn^\mu K_\mu\psi~~, 
\eqno(1)$$
where the Hamiltonian $H$ and the boost operator $K_\mu$ depend also on
the hyper-plane $\sigma$. The Eq.(1) can be split into two unitary 
equations:
$$
\frac{\partial\psi}{\partial a}=-iH\psi~~,
\eqno(2)$$
$$
(\delta_\mu^\nu-n_\mu n^\nu)
\frac{\partial\psi}{\partial n^\nu}=-iK_\mu\psi~~.
\eqno(3)$$
This is standard Dirac theory so far \cite{foot1}. 
BP will retain the second unitary equation (3) while replace 
the first equation (2) by the non-unitary Ito-Schr\"odinger equation of 
standard quantum state diffusion \cite{GisPer}: 
$$
\frac{\partial\psi}{\partial a}=-iH\psi + nonlinear,stochastic~terms.
\eqno(4)$$
BP also present equations for the average state $\rho$, derived
from Eqs.(3,4):
$$
(\delta_\mu^\nu-n_\mu n^\nu)
\frac{\partial\rho}{\partial n^\nu}=-i[K_\mu,\rho]~~,
\eqno(5)$$
$$
\frac{\partial\rho}{\partial a}=-i[H,\rho] + 
L\rho L^\dagger -{\small\frac{1}{2}}L^\dagger L\rho
                -{\small\frac{1}{2}}\rho L^\dagger L~~.
\eqno(6)$$
The first equation is unitary, the second is not.
BP claim to prove that their equations (3,4) as well as (5,6)
are compatible, preserve Lorentz-invariance, and their translation 
invariance can also be pointed out in a restricted sense \cite{foot2}.
Unfortunately, BP are unaware of further basic
features of Dirac's theory. In particular, Dirac theory assures that
various inertial observers have consistent experimental results.
The BP equations fail to do so!

Assume, for instance, that two space-like hyper-planes $\sigma_1$ and
$\sigma_2$ intersect at space time point $x$. Let $A$ be a local scalar
observable at $x$. Then in Dirac's theory the observable $A$ transforms
between $\sigma_1$ and $\sigma_2$ in such a way that its
expectation value will not change:
$$
\langle A\rangle_{\sigma_1}=\langle A\rangle_{\sigma_2}~~,
\eqno(7)$$
where $\langle ...\rangle_\sigma$ stands for expectation values
in quantum states $\psi$ (or $\rho$, in general) taken on the 
hyper-plane $\sigma$. The claimed compatibility and relativistic 
invariance of the BP equations (3,4) and (5,6) are, in themselves, 
irrelevant since the physical consistency (7) of Dirac's theory is lost.
How fatally it is lost the reader shall understand on a 
non-relativistic application. 

{\it The Counter-example.}
We do not even need BP's stochastic equations but the ensemble averaged 
ones (64,65) [cf. my Eqs.(5,6)], together with Eq.(63) which relates 
measured expectation values to the density operator (in the standard way, 
this time). Let us apply these equations to a typical
non-relativistic situation. Consider a free non-relativistic electron
in the authors' reference system $n_0=(1,0,0,0)$. We call the observer who
rests in this reference system {\it R-observer}. Soon we need
a moving {\it M-observer}, too, who rests in the reference
system $n_v=(1,v/c,0,0)$ moving with a {\it small} 
non-relativistic velocity $v$ with respect to the R-observer. 
The electron is non-relativistic for both observers. Assume it remains 
{\it localized} along the right $x$-axis at $x\approx\ell$. Both observers 
will measure the same spin-observable:
$$
A=\vert\uparrow\rangle\langle\downarrow\vert + H.C.~~.
\eqno(8)$$
They will measure {\it in coincidence}! For instance,
the M-observer switches on his apparatus at time $a=0$, in coincidence 
with the R-observer's apparatus at (his/her) time $a_0=\ell v/c^2$. 
We expect that the two measurements lead to the same result in the  
non-relativistic limit $v/c\rightarrow0$:
$$
tr\{A\rho(n_v,0)\}=tr\{A\rho(n_0,a_0)\} +{\cal O}(v/c)~~.
\eqno(9)$$
Note that we can choose an arbitrary large distance $\ell$ so that 
$a_0=\ell v/c^2$ remains relevant (e.g. constant) even for 
$v/c\rightarrow0$. 

For the electron's initial state in the frame of the R-observer 
we choose a superposition of the spin-up spin-down states:
$$
\rho(n_0,0)
=\frac{1}{2}\Bigl(\vert\uparrow\rangle\langle\uparrow\vert 
                 +\vert\downarrow\rangle\langle\downarrow\vert
                 +\vert\uparrow\rangle\langle\downarrow\vert
                 +\vert\downarrow\rangle\langle\uparrow\vert\Bigr)~~.
\eqno(10)$$
(Observe that, initially, the rest observer would measure $1$ for 
the expectation value of $A$.)
We define the following Lindblad generator: 
$$
L(n_0,a) 
\equiv \frac{1}{\tau}\vert\uparrow\rangle\langle\uparrow\vert~~.
\eqno(11)$$
Let the reduction time $\tau$, controlling the strength of the
quantum state diffusion, satisfy the condition $\tau<<a_0$. 
Then the Eq.(64) turns the initial pure state (10) into the mixed one:
$$
\rho(n_0,a_0)\approx
\frac{1}{2}\Bigl(\vert\uparrow\rangle\langle\uparrow\vert 
                 +\vert\downarrow\rangle\langle\downarrow\vert\Bigr).
\eqno(12)$$
For time $a_0=\ell v/c^2$, the quantum state has become reduced
to the mixture of $\vert\uparrow\rangle$ and $\vert\downarrow\rangle$. 
The expectation value of the Hermitian observable $A$ becomes zero for the 
the R-observer's measurement. 

Let's turn to the M-observer. According to the Eq.(65),
he/she initially sees the state
$$
\rho(n_v,0)=
\rho(n_0,0) + {\cal O}(v/c)~~,
\eqno(13)$$
which is identical to the rest observer's initial state up to
terms of the order of $v/c$. So, the M-observer measures
$1$ up to terms ${\cal O}(v/c)$. 

In summary, we can write:
$$
\langle A\rangle_{(n_v,0)}-\langle A\rangle_{(n_0,a_0)}=1+{\cal O}(v/c)
\eqno(14)$$
which, indeed, contradicts \cite{foot3} to the invariance condition (7) 
of the Dirac theory.

{\it Conclusion.}
Abandoning the standard Dirac wave functions has not been the main reason 
for BP's failure. The fatal reason is that relativistic Wiener-processes
do not exist but trivial ones, as is particularly shown in 
the context of continuous wave function reduction theories by, e.g., 
Pearle \cite{Pea} and myself \cite{Dio}. 
Relativistic models of continuous reduction theories should first relax
the Markovian approximation, cf. \cite{Str}.

\bigskip
This work was supported by the EPRSC and by the Hungarian Scientific
Research Fund OTKA-T016047.


\begin{thebibliography}{99}
\bibitem{foot1} 
In standard Dirac theory, $H$ and $K_\mu$ have well-known expressions.
BP are, for unclear reasons, troubling with  their own awkward 
construction. Their wave function (11) [or (21)] is not Dirac's one!
In BP's deliberate representation the Hamiltonian $H$ and the boost 
$K_\mu$ have nothing to do with the well-known forms. One has no guess
why BP depart from Dirac's standard wave functions.
\bibitem{foot2} That translation invariance would be broken by the 
author's dynamic equations we demonstrate easily. For instance, let the 
parameters of three space-like hyper-planes
$\sigma_1,\sigma_2,\sigma_3$ be, in obvious notations, chosen as follows: 
$n_1=n_2=(1,0,0,0)$, $n_3=(2/\sqrt{3},0,0,1/\sqrt{3})$, $a_1=1$, $a_2=2$, 
$a_3=2/\sqrt{3}$. Let the two other reference frames $R_1,R_2$ have
$(0,0,0,\sqrt{3}-2)$ and $(0,0,0,2\sqrt{3}-2)$ as their shifted origins. 
Elementary calculations show that  
$a_1=a_3$ in $R_1$ and $a_2=a_3$ in $R_2$, while $n_1,n_2,n_3$ do not
change. Consider the averaged states
$\rho_1,\rho_2,\rho_3$ respectively on the three hyper-planes.
We apply Eq.(65) of BP in frame $R_1$ to relate 
$\rho_1$ and $\rho_3$: they turn out to be unitary equivalent. Similarly, 
we apply Eq.(65) in frame $R_2$ to $\rho_2$ and $\rho_3$: they, two, will
be unitary equivalent. This implies that $\rho_1$ and $\rho_2$ must be
unitary equivalent. But these latters are related non-unitarily by Eq.(64)
(no matter which inertial frame is chosen).
\bibitem{foot3} $\ell$ or $1/\tau$ (or both) diverge in the limit 
$v/c\rightarrow0$. This does however not invalidate our Counter-example.
First, it makes sense to assign any finite values to $\ell$ and $1/\tau$.
Second, the violation of Eq.(7) by Eq.(14) is obvious already at finite 
$v/c$, i.e. at finite $\ell$ and $1/\tau$.
\end{thebibliography}
\end{document}